# Accurate computational evolution of proteins and its dependence on deep learning

Prabha Sankara Narayanan[1], Ashish Runthala[2]*

1. Department of Computational Biology and Bioinformatics, University of Kerala, India

2. Department of Biotechnology, Koneru Lakshmaiah Education Foundation, India

Email: *ashish.runthala@gmail.com

**Abstract**

Enzyme is the major workhorse to carry out the diverse cellular functions. It catalyzes the biological reactions with a high specificity, with its topology playing a crucial role. For ecologically safe production of numerous bioproducts including drugs and chemicals, we have been striving to design the industrially useful enzyme molecules with highly improved catalytic capability. As the sequence space is enormous for an enzyme, its quick and effective exploration is quite improbable for the mutagenesis studies whose accuracy is greatly reliant on the prior information of the mutated sites and the extent of rigorous screening of the mutant libraries. Although directed evolution methods significantly aid the construction of a functionally improved molecule, their credibility depends on the successful excavation of the functionally similar sequence space in the available databases, encompassing billions of proteins. As deep learning methods aid us to extensively uncover the underlying network of all the key catalytic positions without any experimental data, their implementation has reliably increased the accuracy of directed evolution. The chapter comprehensively explains data mining and deep learning methods to further showcase their importance in enzyme engineering methods. The key biological and algorithmic limitations of these deep learning methodologies are lastly highlighted.

## 1. Introduction:

Cellular system has naturally evolved all the protein sequences to govern their molecular networks at the required rate under the constrained microenvironment, with enzymes or ~4% proteins being the key regulators [1]. Enzymes naturally evolve in all organisms, and absorb the mutations to constantly evolve and tackle the environmental constraints in a better way. Their catalytic activity is remarkably higher than all chemical means, unbeaten till date [2]. Their importance is quite clear from the fact that the market cap for biotechnology products is increasing at an annual rate of 9.2%, and is estimated to reach $950 billion from current $497 billion by 2027 (https://www.gminsights.com/industry-analysis/biotechnology-market). Responding to these estimates, the leading industrial firms including Japan (Japan Bioindustry Association, Biotechnology Strategy Council), Europe (European Technology Platform for Sustainable Chemistry) and USA (Advanced Technology Platform, National Institute of Standards and Technology) have realized the worth of biotechnology for cost-effective and sustainable growth, as per McKinsey's report of 2020. Although the big firms are leveraging the biotechnology strategies into their developmental pipelines, the biological option is only prioritized when the chemical arsenal has badly failed to produce a biomolecule at a high purity in a cost-effective manner.

The industrial requirement of several key methodologies for (a) screening the most potent enzyme source for producing a required molecule, (b) altering the metabolic pathway of the host cell to boost the yield, (c) customizing the active-site to tune-up the enzyme for a non-natural substrate, (d) engineering the reaction mechanism with alternative molecules for overcoming the rate-limiting steps, has significantly stimulated the industrial developments, as briefly detailed in the following Figure1. While the first two strategies aid the construction of an ideal and highly fruitful host cell for the expression of a target protein and get its high yield, the latter ones help in engineering the catalytic activity of the enzyme for improving its yield. Here, enzyme engineering and directed evolution are at the core of the entire protocol, and majorly regulate the overall cellular yield. The directed evolution methodology for engineering the enzyme structure has thus been recently pioneered by Prof. Frances Hamilton Arnold in 2018, for which she won the nobel prize in chemistry.

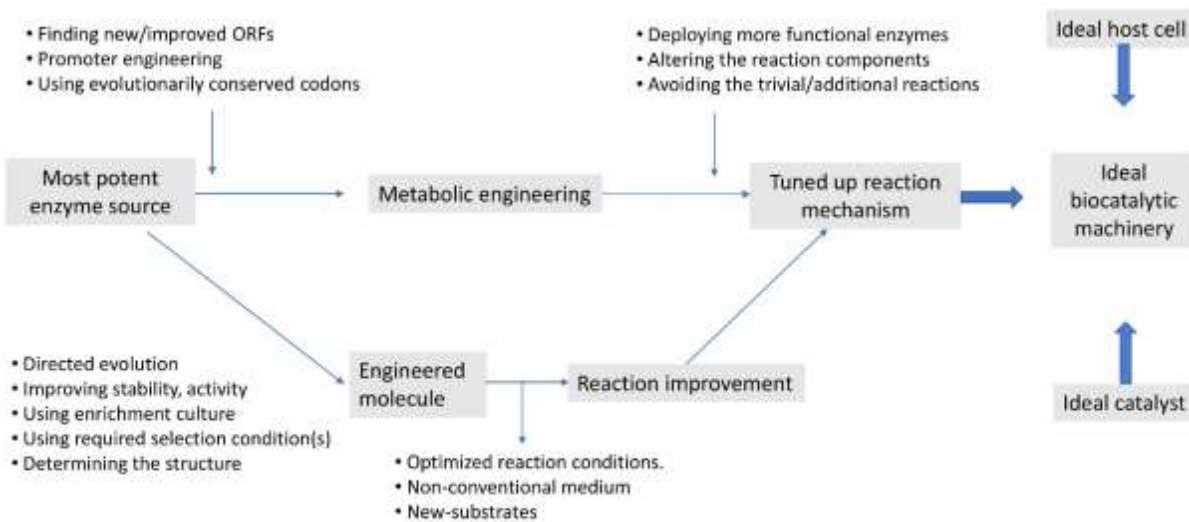

Figure 1: Major strategies for engineering the enzyme and host system to design an industrially useful host system. Importance of both the improved host cell system and the catalytically efficient enzyme is illustrated.

To meet demands, the enzyme activity needs to be improved to produce diverse categories of biomolecules at high enantiomer efficiency and purity, and hence being most important amongst all these strategies, the directed evolution methodology majorly encapsulates the steps required to design an industrially useful enzyme. However, to accurately engineer an enzyme molecule, the most active enzyme resource should be reliably screened, as shown in Figure1. Neither a wrongly selected enzyme can be evolved to attain the desired results, nor a rightly selected enzyme can be evolved if the evolutionarily favored residues at the key positions are not efficiently revealed. It thus becomes crucial to excavate the sequence datasets for extracting all sequence homologs for the target enzyme, and here, the deep learning methodology plays a crucial role. The article explains the directed evolution approach along with its major steps, then goes on to explain the underlying notion of data mining and deep learning approaches. Major computational and biological limitations of this deep-learning based enzyme engineering methodology are lastly explained to unleash the future avenues for further research.

## 2. Directed evolution

Although the naturally available enzyme molecules are the best catalysts, their tolerance to organic solvents, lack of stability and limited scope of substrates make them unsuitable for several industrial processes. Ever since 1976, enzymes have been mutated through different mutagens like X-rays, as an attempt to develop a better strain with an improved phenotype [3-5]. Although this method has been used for developing the strains to produce a large quantity of the desired product molecules and to introduce a novel biosynthetic pathway [6-8], it is ineffective and does not promisingly incorporate the desired change, making it useful for the organisms with a short replication cycle. With the advent of a generic site-directed mutagenesis in 1978, mutations at certain key positions have been introduced in the enzymes to excavate their detailed functional mechanisms. As the effect of a mutation on the activity is often unpredictable, this rational enzyme engineering methodology is not always successful. Thus, the site-saturation mutagenesis has been extensively used to introduce the functional change, and uncover its effect on the catalytic properties.Besides increasing the thermostability of glucose isomerase [9] and improving the stability of subtilisin [10], directed evolution method has led to the construction of enzyme variants with an improved activity for unconventional substrates [11], pH stability [12], altered enantioselectivity [13]. It has hence become one of the powerful strategies of biotechnology, and has suggested that the functionally beneficial mutations are localized at the unexpected parts of an enzyme, hinting why rational methodologies have struggled to mutate the key sites [10, 14].

Directed evolution theoretically revolves around the designing of a mutant library, separation and cloning of individual mutants, expression of mutated genes for evaluating their phenotype, scrutinization and identification of functionally best genotype, with iterative reoptimization and scrutiny. For constructing a functionally improved enzyme molecule than the wildtype, this strategy utilizes the following key steps, and the last three steps are often iteratively deployed to attain the best results.

1) *Selection of the most potent enzyme for meeting the desired objective*

    It is the most crucial step of the entire strategy. As the functionally similar sequences have already been preselected by natural evolution, their consideration allows us to easily screen the residue variations at all the functionally crucial positions, this step is indispensable for the construction of a smart mutation library for the desired enzyme. Through the sequence

homology at different positions of the target protein, it promisingly captures the most evolutionarily favored/variant subsequences.

2) *Construction of the library of mutants*

   As it is impossible to randomize every position of a protein, usually encoding 200-300 amino acids, the mutant library is designed on the basis of only a few specific positions. Besides allowing us to evaluate the effect of a mutation at the designated sequence sites, causing the construction of a small number of mutants, it aids us to easily assess the combinatorial effect of several mutations together. The sophisticated computational methodologies of protein modelling, docking and molecular dynamic simulations have allowed us to restrict the mutagenesis only at key sites. Recent advances in the understanding of protein structure and precise estimates of its functional dynamics have also empowered us to reliably select the most crucial target residues for site-saturation mutagenesis [15].

3) *Fixation of the selection methodology that leads to scrutinization of the numerous clones for finding the most potent ones*

   Though the selection marker is easy to choose, scrutinization of thousands of different mutant clones poses a major challenge to a directed evolution experiment. The screening of most active clones is not straightforward, and this problem is tackled by developing the high throughput screening methods, and reducing the size of the mutant library, focusing it on the key catalytic positions.

4) *Re-diversification of the prioritized candidate mutant sequences*

   After the initial filtering of the most beneficial mutations, the top-ranked mutant is diversified by introducing mutations at the other designated positions. It aids the assessment of combinatorial diversity of mutations, though there is a high chance to miss or neglect the more active mutant. A popular approach deployed for this purpose is the combinatorial active site testing (CASTing), wherein several residues lining the active site are simultaneously mutated to evaluate their synergistic effect on the overall catalytic activity [16]. Several strategies like SCHEMA have been developed for this step to assess the recombinations that most likely disrupt its overall fold [17] and to thus reduce the number of combinatorial mutants.

5) *Fixation of more stringent selection methodology*

As the mutants now have multiple mutations at the probable beneficial sites, the selection of the best mutant is now done through more stringent conditions.

For finding the most beneficial mutations, the complete structural analysis of a protein is impossible due to its astronomical sequence space, depending on its sequence length. A variety of methodologies have been developed for these steps [18-22], and their detailed discussion is beyond the scope of this article. To robustly cover the sequence space of a protein sequence, random mutagenesis is used and then after finding the hotspots, targeted mutagenesis is deployed to find the most beneficial mutations [23]. Thus, localizing the target positions for mutagenesis is an important step for scrutinizing the most plausible sequence space of a protein. In a nutshell, using the promiscuous nature of an enzyme, the directed evolution method is used to carry out the non-natural reactions at a significantly high catalytic rate (Figure2). From the promiscuous region of protein residues, responsible for interacting with a substrate, the directed evolution methodology inculcates the desired variations in a stepwise iterative manner. Proteins evolve under strong selection pressures, according to studies, and functionally neutral mutations are favourable for further adaptation [24].

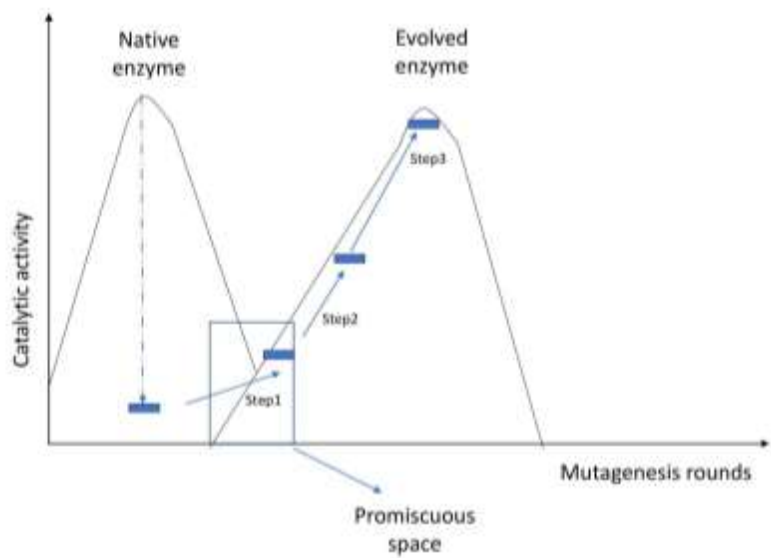

Figure 2: Generic strategy of an iterative mutagenesis based directed evolution strategy. The stepwise construction of a more evolved enzyme variant is shown, initiating from the promiscuous sequence space.

3. **Data mining**

The units of information produced from various sources are collectively termed as data, and data mining is the generic term given to the exploration of huge chunks of data (*Big data*) by finding patterns to understand, analyze and interpret the data. Data mining techniques are generally used to manipulate the available data and extract the required information, and accordingly the data warehousing strategies have changed to easily extract data from gigantic information resources. Data mining plays a vital role in organizations as it narrows down the ambiguity in decision making through profound data analysis. Standardized datasets are built by implementing data mining and are further applied by machine learning algorithms for predicting outcomes of a given problem [25, 26]. Data mining processes the raw data to obtain a structured dataset to unleash an underlying insight of the given problem. The methodology is widespread across different areas of biology including protein structure prediction [25, 26] and medicine [27-30], and is now making its mark in the directed evolution and enzyme engineering field as well.

Data scientists generally perform exploratory data analysis and apply machine learning algorithms to discover correlations and patterns to describe data [27, 29]. Data mining involves a few key steps, viz. understanding the objectives, data collection and preparation, data cleaning, data integration, data reduction, data transformation and finally evaluating the results obtained along with knowledge representation [28, 30]. The initial step of identifying relevant objectives is crucial that signifies the grounds for the data gathered and defines a tentative boundary of the estimated data. It involves data cleaning, deduplication, dimensionality reduction, noise reduction whilst the outlier data can notably shoot up the computational overhead that causes hindrance in efficient training. Once the objective is clear, preparing the data becomes effortless and viable, the outlier data is easily purged followed by machine learning or deep learning model development for identification of patterns in the given dataset through statistical formulas, rules of association and other statistically relevant scoring parameters [28, 30].

4. **Biological data mining**

The data mining techniques have now been significantly used in different research sectors of biology to get molecular insights from the enormous Omics and clinical data. The big data in biology has crossed over petabytes (PB) and exabytes (EB) and it has led to the development of several strategic solutions for progressive analysis and interpretation. Data mining is applied in two ways: descriptive data mining that is used to find interesting patterns to describe data and predictive data mining that predicts the behavior of the model based on the given dataset. While the former is a knowledge-based predictive method and is used to develop the problem-specific solutions, the latter one represents a 'black box' that predicts outcomes that are just some values with no interpretation. However, biological research requires a combination of both predictive and descriptive data mining methodology to predict outcomes representing knowledge and interpretation to unveil state-of-the-art problems [31]. To solve all such problems, deep learning has been playing a vital role to produce accurate results and predictions through advanced automation and modelling strategies [32]. It can be defined as an iterative approach of learning to extract relevant features from a given dataset for predicting certain outcomes.

Biomedical research has been using the data mining strategy through deep learning in a variety of problems [33]. Prediction of peptide properties [34], including tandem mass spectra [35], ion mobility [36], retention time [37], peptide identification [38], protein inference [39] and peak detection [40] are some of the key examples in biomedical research that involves deep learning. Though deep learning models showcase remarkable results, certain factors or parameters in its application require detailed knowledge and domain expertise.

## 5. Deep learning

Deep learning is a technique of deep feature extraction from the data, beyond the manual exploration. A Deep learning model is a network of interconnected artificial neurons, termed as an Artificial Neural Network (ANN). An artificial neuron is a mathematical representation of a biological neuron. Besides deploying ANN, it involves an input layer, several intermediate layers and an output layer that predicts the categorized set of resultant data. The depth and complexity of the network is defined by the number of intermediate layers. When an input is passed to the input layer it gets multiplied with weights and added up with biases. While a statistical weight is a coefficient that determines how important is an input value to further pass it onto the hidden layers, a bias defines a constant value that sets the offset of the output value by making it either positive or negative. When the weights are statistically insignificant, their corresponding nodes are eliminated, and that process is known as dropout. Model accuracy can be improved if the input data is preprocessed/ cleaned and standardized that is amenable to the models, and is then fed to the ANN. A neural network also consists of activation functions that transform the weighted sum of input data to produce an output that is either linear (real number values) or nonlinear (0 or 1). Classification and clustering are the usually used methods for predicting the most probable outcomes of a given dataset. While the classification method is a supervised learning technique with labelled data as input, the clustering method comes under unsupervised learning technique with unlabeled or unstructured data as input. The dataset is divided into training and test sets. Model evaluation is performed using a validation set (20% training data). Observing a significant deviation between the resultant and expected result, leading to a greater loss, the deep-learning model is tuned through variations in hyperparameters and train-test data split to finally get an unbiased estimate of the final trained model. The test set is the new or unknown data that is fed as input to the developed model for comparing the differences between the output and the desired data. To effectively train a neural network, enormous amounts of data is required to capture the minute features that are representative. Models trained with larger datasets can be utilized further for predicting similar kinds of datasets which is known as transfer learning. The concept of transfer learning is used to transfer the patterns obtained from model training to the similarly structured datasets for predicting outcomes [41].

Model tuning is an approach to increase the model accuracy by tuning the hyperparameters. The key hyperparameters that affect the accuracy of a neural network include Epoch, learning rate, optimization algorithm and batch size. Epochs are the number of iterations by a model to train itself with the complete dataset (training data). Learning rate is an important parameter that determines how a model can arrive at its best accuracy. Optimization algorithms are essential to minimize the loss obtained from training a model. Loss function is a measure of error rate that signifies the accuracy of resultant data against the expected outcome and is used as a feedback signal through optimization method. The method of feedback using optimal weights is known as backpropagation and is aimed to minimize the error rate by adjusting the weights to make an optimum model [42]. To bring the loss to an optimal minimum, several optimizers viz. gradient descent, stochastic gradient descent (SGD), mini batch stochastic gradient descent (MB-SGD), adaptive gradient (AdaGrad), nesterov accelerated gradient (NAG), Adam and root mean square prop (RMSprop) are normally used. Batch size is another important parameter that takes input data in batches for training to optimally utilise the available computational resource. For instance, if the data size is 1 lakh samples for training then it can be split to a batch size of 10,000 samples to be trained at a time and likewise 10 batches of the whole dataset [43].

Protein engineering has become a predominant step to divulge its structure and function. In the recent past, the directed evolution algorithms have started using the deep learning methodology to improve its key catalytic property, as discussed earlier. Directed evolution requires at least a single functional parent and a sequence function that is locally smooth [44]. Without the prior knowledge of any other modelling pathways, the deep learning method uses the detailed sequence profile of a target sequence. For the aligned sequences, it investigates the most conserved/ variant position for ranking the alternative residues at every position of the target sequence to decipher the most prevalent residues, preselected by nature through natural evolution. This approach thus adopts the process of natural selection from the Darwinian theory that represents "survival of the fittest". It maps the sequence to its function and the undertaken catalytic, structural properties to generate correlation data. As the evolutionarily preferred mutational data is used for the input profile, it spans the biologically favored region(s) of the sequence space for a target sequence and has thus an increased likelihood of building a more active sequence. New variant identification becomes a rational procedure through this method making it viable to select the best fit [24]. While the manual strategy is only restricted to the study of most active mutants to effectively map the sequence space of a protein, the deep learning methods also use the information from the discarded variants to decipher the characteristic features that more closely fits the function of an evolved protein to its function. This strategy has thus shown the potential to accelerate protein engineering [24], and thus several research groups have used it for varied objectives/ problems.

For extracting the underlying functionally crucial features of a protein, Ethan's group has deployed deep learning methodology in their methodologies for explaining the structural, biophysical and evolutionary relationships within the protein sequences [45]. For exploring the hidden areas in the fitness landscape of a protein, several research groups have used this strategy. One exemplary research work has improved the prediction accuracy of a single point mutation over the other simple machine-learning models [45]. This work uses the Unirep database, encompassing more than 20 million proteins to extract the fundamental features of the proteins through global unsupervised pre-training [45]. It then uses the unsupervised weights to reliably train its deep learning model over the sequences, evolutionarily- related to target protein, calling this step as evotuning. This method combines the features obtained from both global and local sequence

landscapes. It has specifically quoted the following core findings related to the deep learning methodology:

   a. Vectorization method of amino acids captures the physicochemical properties of each amino acid group.
   b. Vectors, representing the proteome of a single organism, encapsulates the taxonomy data.
   c. The internal hidden states can be reliably examined to investigate the statistical correlation between the input and output data, for every specific feature.
   d. Deep-learning method significantly improves the prediction accuracy of a single point mutation over other machine-learning based models.

## 6. Current research status

The machine learning strategies have been remarkably successful in the field of protein engineering and have been extensively used in the recent past for targeting the key objectives (Table1). Traditional directed evolution methods have been judiciously tweaked to deploy these computational strategies for getting the reliable outcomes, as exemplified below for the tabulated enzymes.

| # | Enzyme | Objective of the study | Reference |
|---|--------|------------------------|-----------|
| 1 | Green fluorescent protein | Decode the phenotype of distantly related functional variants, and build virtual fitness landscape | 45 (2019) |
| 2 | Halohydrin dehalogenase | Improve the catalytic function | 46 (2007) |
| 3 | Phosphomannose isomerase, TEM-1β-lactamase | Find the gain-of-function mutations | 47 (2020) |
| 4 | Toluene-4-monooxygenaseKR1 | Improve the enzyme activity | 48(2010) |
| 5. | CYP2D6 | Prediction of metabolic activity of CYP2D6 genotype and functional classification of haplotypes | 49 (2019) |
| 6 | Histone Reader | Identification of strong binders | 50 (2020) |
| 7 | Surfactin phosphopantetheinyl transferase, Acyl carrier synthase | Design de novo peptide substrate | 51 (2018) |
| 8 | Epoxide Hydroxylase | Improving the enantioselectivity | 52 (2018) |
| 9 | Nitric oxide dioxygenase | Improve enantiomeric selectivity | 53 (2019) |
| 10 | Aldehyde dehydrogenase, 1-deoxy-D-xylulose-5-phosphate synthase | Increase the enzyme solubility | 54 (2020) |
| 11 | Deoxyribose-5-phosphate aldolase | Modify substrate promiscuity | 55 (2020) |
| 12 | Unique point mutations from the Protherm database | Improve the accuracy of stability prediction methods | 56 (2020) |
| 13 | *Bacillus thermoleovorans* pullulanase | Increase the thermostability | 57 (2020) |
| 14 | Phosphotyrosine | Identification of novel SH2-superbinders | 58 (2020) |

| 15 | HER 2:- Human Epidermal Growth factor receptor 2 | Predict HER 2-specific subset | 59 (2021) |

a. **Improve the catalytic function of the enzyme [46, 47]:** ProSAR is a recombination-based directed evolution which involves statistical analysis of protein sequence activity relationships. The combination expedites mutation-oriented enzyme optimization by capturing information in the sequence-activity data. Thus, the strategy identifies beneficial variants as well as mutants with reduced function, and constructs the evolved variant of halohydrin dehalogenase, with a volumetric productivity of ~4,000-fold [46]. As an alternative for the energetic-approaches, the methodologies deploy a 3D convolutional neural network as a guide to connect amino acids with their neighboring chemical microenvironment for identifying the novel gain-of-function mutations. As recently shown [47], it is proven to identify the chemical interactions responsible for the gain-of-function phenotypes for various mutations.

b. **Improve the enzyme activity [48-50]:** For an increased production of a potent antioxidant, hydroxytyrosol, directed evolution and rational design methodologies have been recently integrated for improving the activity of toluene-4-monooxygenase against its substrate 2-phenylethanol [48]. The computational learning strategy has hereby played a vital role in constructing a statistical model to evaluate the sequence-fitness function for selecting a minimal count of 16 best mutants from the pool of ~13000 variants. Machine learning has played a phenomenal role in this strategy, and the resultant best variant has exhibited a 190-fold higher activity than the that of the wild type.

A deep learning model has also been deployed for the functional annotation of Cytochrome P450 2D6 haplotype, as the pace of functional assignment is far lower than the sequencing throughputs [49]. With an accuracy of 88% on the training dataset, the strategy could be a fruitful tool for attributing functional details to functionally uncurated haplotypes. An ordinal regression has also been used to extract the reliable predictive features for the estimation of binding affinity of chromodomain protein CBX1 mutants towards the histone peptide H3K9me3. On basis of the peptide microarray binding experiments, the study uses the nine sites of CBX1 and through the predictive features, it extracts the three top ranked mutations for each position to reduce the search space from the possible set of $20^9$ sequences to $3^9$ (19683) entries. With minimal monetary and experimental resources, the study leads to the construction of strong binders with a dissociation constant, quite close to the best variant constructed through the microarray studies [50].

c. **Design de novo peptide substrate [51]:** The machine learning strategy has recently been utilized to simultaneously optimize the biochemical features of peptides and construct the potential peptide substrates for 4'- phosphopantetheinyl transferase. As it assists the selection of potential peptide substrates from the experimentally known dataset through mathematical scores, it proves its efficacy in leveraging the construction of peptides for several biological functions that are not easily accessible to experimental methodologies like phage-display.

d. **Improve the enantioselectivity [52, 53]:** By estimating the functional fitness of mutants for the candidate protein epoxide hydrolase of *Aspergillus niger*, the machine learning model, trained through more than 500 physicochemical and biochemical properties of amino acids, has been successful in designing mutants with an improved enantioselectivity [52]. Frederic Cadet's group has developed this innovative sequence activity relationship (Innov'SAR) method by combining the experimental and computational methodologies. Permuting 9 single point mutations across the active-site, $2^9$ (512) mutants are generated to screen the best variant, and thus this method can be iteratively used to guide the iterative saturation methodology.

Simultaneous exploration of the protein sequence space encoded by several amino acids could be an unachievable task experimentally, and hence machine learning has been significantly more useful than the usual directed evolution experiments [53]. Testing on the empirically known fitness landscape benchmark dataset of human GB1 binding protein, the trained machine learning-guided directed evolution method has predicted variants with higher fitness than the conventionally evolved variants. The algorithm successfully constructs an improved variant of the putative nitric oxide dioxygenase of *Rhodothermus marinus* with 93% ee (enantiomeric excess) [53].

e. **Increase the solubility of enzyme [54]:** For improving the protein solubility through computational mutagenesis, the 20/30-residue sequence tags are added to six different test proteins (1-deoxy-D-xylulose-5-phosphate synthase, valencene synthase, chalcone synthase, alcohol dehydrogenase, tyrosine ammonia-lyase and 4-coumarate-CoA ligase)

[54]. Deploying a support vector regression model, it aims at designing the short peptide tags for increasing the solubility of target enzymes. With an iterative crossover and mutagenesis of parental sequence tags, the offspring segments are created and the ones with higher solubility are further used as the parent tags to evolve them towards more soluble variants. The machine-learning based optimized sequence tags significantly improves the solubility and activity of test proteins and proves it as a beneficial tool to improve the performance of an enzyme.

f. **Modify substrate promiscuity [55]:** *E.coli* 2-deoxy-D-ribose 5-phosphate aldolase is engineered to induce a preferential affinity towards smaller non-phosphorylated aldehydes like acetaldehyde [55]. The 24 residues within/close to the active site are considered to generate the variants with 1-3 mutations, for further testing their preferential catalytic nature against three different substrates. The mutants are studied by machine learning through Gaussian processes, and for third mutagenesis, a set of 48,000 mutants are created to select the top-18 mutants with the highest predicted acetaldehyde specificity for experimental characterization. Machine learning has thus been very fruitful in screening the out-of-the-box mutants to select the natively best ones, with a high preferential bias for the desired substrates.

g. **Identify structural preferences of increased stability [56, 57]:** As accurate assessment of mutations on the stability of proteins is key to functionally understand their catalytic mechanism and improve their stability, a set of 21 usually deployed protein stability prediction algorithms, also including the machine-learning strategies, are analyzed to evaluate their stabilizing effect against the experimentally benchmarked dataset [56]. Although the statistical scoring-based machine learning methodology demonstrates the best prediction, its accuracy is usually skewed towards the highly destabilizing mutations, still needing further refinements. For an increased credibility of machine learning based methods, several such tools, i.e. I-Mutant 3.0, dFire and FoldX) have even been simultaneously used to improve the thermostability of thermophilic pullulanase of *Bacillus thermoleovorans* [57]. Through the detailed molecular dynamics simulation of this modelled protein, the study validates 6 of the 17 highly stable mutants at non-conserved and non-catalytic sites. However, the study still needs to yield the prediction accuracy of a double/triple/multiple mutant, usually constructed through the top-ranked mutations.

h. **Identification of novel SH2-superbinders [58]:** Randomizing 8 variable residues within the phosphotyrosine(pTyr)-binding pocket of the SH2 domain of human pp59fyn (Fyn), the sequence space of the variant library is restricted with an aim to design novel variants with an increased binding affinity to pTyr peptides. Quantifying the affinity through fluorescent polarization, a triple mutant of SH2-domain with a significantly higher binding affinity is constructed. The machine learning-guided study shows a novel way to identify the pTyr-containing proteins from tissues under divergent physiological/ pathophysiological conditions.

i. **Predict HER 2-specific subset [59]:** As an attempt to minimize the complexity of the lead optimization methodology of drug discovery and increase the therapeutic potential of antibodies, the deep learning model has been recently used for simultaneously optimizing several parameters including solubility, viscosity, pharmacokinetics, immunogenicity, and expression level. Screening the binding specificity of an experimental library of $5 \times 10^4$ therapeutic antibody trastuzumab variants towards the human epidermal growth factor receptor 2 (HER2), the trained neural network is used to construct numerous lead candidates. Experimental evaluation of 30 such randomly selected variants validates their improved antigen specificity, and proves that deep learning could be a reliable methodology for engineering the antibody structure.

**7. Major limitations**

a. **Methodological:** As a deep learning model requires training on large datasets, the major challenge in implementing it for every biological problem would be the unavailability of substantially larger datasets. In case of enzyme engineering, the initial training data is collected from the heterogeneous databases, as discussed in the guidelines provided by the standards for reporting enzyme data (STRENDA) [60]. Although the public repositories are kept open-access for datasets, the available data is mostly unstructured and this poses a big challenge to construct algorithms that reliably extract the data. Another issue that needs to be addressed is the utilization of minimal data for model training and considering unstructured or unlabeled data for training. Moreover, the unsupervised learning approach for pattern recognition is not widely explored in the field of protein engineering. Usually, the uncleaned data is used to train the machine learning model and thus the generated outcomes do not have any context to be interpreted.

Data collected from directed evolution studies consists of both functional and non-functional sequences. Machine learning methods normally discard the nonfunctional entries during preprocessing and it leads to the loss of useful training information. Domain expertise is an important factor in terms of handling protein related data. Expert opinion aids in purging the trivial data and predicting the desired outcomes, after a detailed benchmarking that is not always implementable.

b. **Biological:** Machine learning has invoked several challenges in many different aspects as it is predominant to recognize the kind of data and the approaches to be applied. For the directed evolution of proteins, the functionally similar dataset should first be excavated, and hence, the objective may not be accomplished if the minimally required sequence dataset is not complete or if it encompasses additional functionally different entries. Being mesmerized by the continuous increase in the sequence databases, biologists have hoped that evolutionary incongruence, observed with a few entries, would be resolved with sufficiently larger datasets [61]. Contrary to these beliefs, the phylogenetic reality, or the accuracy of residue substitutions across a functional protein family has been shown to be much more complicated [62, 63]. Nevertheless, for a directed evolution experiment, the congruence of sequence overlaps has always been strived within the sequence datasets to extract the biologically feasible alternative residues at all the key positions. Hence, it is important to filter the reliable features for classifying the input data and for improving the accuracy of final outcomes, and hence, detailed literature knowledge becomes mandatory to obtain reliable results. Sufficiently large sequence dataset is thus often used along with the traditional protein analysis methods like phylogeny to do a preliminary classification for further guiding the machine learning protocol to attain the required data. If many functionally diverse sequences are randomly collected to build the initial dataset for tuning the machine learning model, there is a high risk of getting uninterpretable and meaningless false positive results. As the lack of several informative details viz. standard protein ontology, sequence strength, length and diversity and the origin and strength of heterogeneous data usually complicates the entire training model; scrutiny of the first dataset should be carefully done before initial training. A clear objective solves this problem to some extent and allows the integration of several biological databases for reliably increasing the sample size. The current state-of-the-art research should thus be made available as a separate database to avoid ambiguity data, usually collected while studying the available literature [64]. Lastly, the complicated computational biology problems should be presented on a public forum where researchers can freely exchange their ideas and collaborate to solve the biological riddles.

**8. Conclusion:**

Computational learning based enzymatic evolution is a multi-objective optimization problem. With an ever-increasing number of protein sequences and structures, the sequence space of a protein sequence should be swiftly explored to excavate all possible alternative residues at each potential position for constructing the best possible functionally evolved sequence. To evolve the functionally improved enzymes, deep learning is now being extensively used, and its accuracy is dependent on two sub-problems: (A) completeness and correctness of the biologically correct sequence profile, and (B) biological feasibility of every screened mutation at its target site within a protein sequence. While the former problem could be reliably solved through more rigorous sequence alignment and statistical methods, the latter one is computationally infeasible to estimate. Iterative tailoring of the algorithmic coefficients and key parameters is the biggest strength of a deep learning method that empowers it to learn from both the insignificant variants and allow it to simultaneously optimize multiple properties and predict more promising sequence variants. Deep learning methods should thus be integrated with the high-throughput methods to map the sequence space efficiently and quickly for constructing the improved variants for various pharmaceutical and industrial applications.